\documentclass[aps,twocolumn,floats,pre]{revtex4}
\usepackage{graphics,graphicx,epsfig}
\usepackage{amssymb,color}
\usepackage{epsf,epstopdf,wrapfig}
\usepackage {amsmath}

\newcommand{\beq}{\begin{equation}}
\newcommand{\eeq}{\end{equation}}
\newcommand{\beqn}{\begin{eqnarray}}
\newcommand{\eeqn}{\end{eqnarray}}

\begin{document}

\title{Morphogenesis at criticality?}

\author{Dmitry Krotov,$^{a,b}$ Julien O. Dubuis,$^{a,b,c}$  Thomas Gregor,$^{a,b}$  and William Bialek$^{a,b}$}

\affiliation{$^a$Joseph Henry Laboratories of Physics, $^b$Lewis--Sigler Institute for Integrative Genomics, and $^c$Howard Hughes Medical Institute, Princeton University,
Princeton, New Jersey 08544, USA}

\begin{abstract}
Spatial patterns in the early fruit fly embryo emerge from a network of  interactions among transcription factors, the gap genes, driven by maternal inputs.  Such networks can exhibit many qualitatively different behaviors, separated by critical surfaces.  At criticality, we should observe strong correlations in the fluctuations of different genes around their mean expression levels, a slowing of the dynamics along some but not all directions in the space of possible expression levels,  correlations of expression fluctuations over long distances in the embryo, and departures from a Gaussian distribution of these fluctuations.  Analysis of recent experiments on the gap genes shows that all these signatures are observed, and that the different signatures are related  in ways predicted by theory.  While there might be other explanations for these individual phenomena, the confluence of evidence   suggests that this genetic network  is tuned to criticality.  
\end{abstract}

\date{September 10, 2013}

\maketitle

Genetic regulatory networks are described by many parameters: the rate constants for binding and unbinding of transcription factors to their target sites along the genome, the interactions between these binding events and the rate of transcription, the lifetimes of mRNA and protein molecules, and more.  Even with just two genes, each encoding a transcription factor that represses the other, changing parameters allows for several qualitatively different behaviors  \cite{hasty+al_01}.   With delays (e.g., in translation from mRNA to protein), mutual repression can lead to persistent oscillations.  Alternatively, if  mutual repression is sufficiently strong, the two genes can form a bistable switch, admitting both on/off and off/on states, with the choice between these states  modulated by  inputs to the network \cite{toggle}.  Finally, if  interactions are  weak, the two interacting genes have just one stable state, and the  expression levels in this state are controlled primarily by the inputs.   The bistable switch and the graded response to  inputs are limiting cases; surely the truth lies somewhere in between.  But if we imagine smooth changes in the strength of the repressive interactions, the transition from graded response to switch--like behavior is {\em not} smooth:  the behavior is qualitatively different depending on whether the relevant interactions are stronger or weaker than a critical value.     Here we explore the possibility that the gap gene network in the {\em Drosophila} embryo might be tuned to such a critical point.

\begin{figure}[b]
\includegraphics[width=\linewidth]{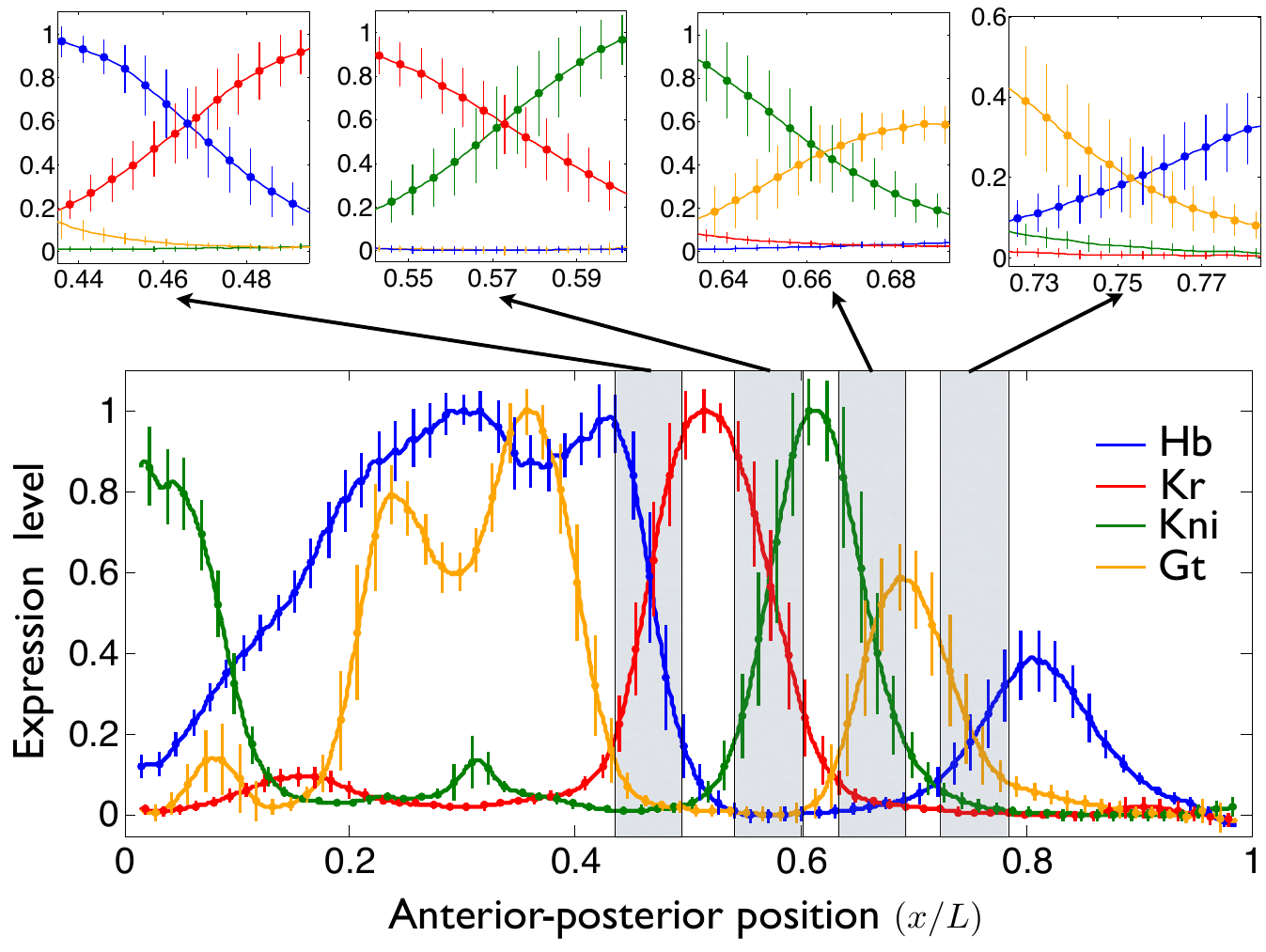}
\caption{Normalized gap gene expression levels in the early {\em Drosophila} embryo, from Ref \cite{expts}.   Measurements by simultaneous immunoflourescent staining of all four proteins, along the dorsal edge of the mid--saggital plane of the embryo, 38--49 min into nuclear cycle 14; error bars are standard deviations across $N=24$ embryos.   Upper left shows an expanded view of the shaded regions, near the crossings between Hb and Kr levels, where just these two genes have significant expression, and similarly for the Kr--Kni, Kni--Gt, and Gt--Hb crossings in upper panels from left to right. \label{fig1}}
\end{figure}

Early events in the fruit fly embryo provide an experimentally accessible example of many questions about genetic networks \cite{dev_general, lawrence_92, gerhardt+kirschner_97}.  Along the anterior--posterior axis, for example, information about the position of nuclei flows from primary maternal morophogens to the gap genes, shown in Fig \ref{fig1} \cite{jaeger_11,expts,bits},  to the pair rule and segment polarity genes.  Although the structure of the gap gene network is not completely known, there is considerable evidence that  the transcription factors encoded by these genes are mutually repressive \cite{jaeger_11,jackle+al_86,treisman+desplan_89,hoch+al_91}.  If we focus on a small region near the midpoint of the embryo (near $x/L = 0.47$), then just two gap genes, {\em  hunchback} (Hb) and {\em kr\"uppel} (Kr), are expressed at significant levels, and this is repeated at a succession of crossing points or expression boundaries: Hb--Kr, Kr--Kni ({\em knirps}; $x/L = 0.57$), Kni--Gt ({\em giant}; $x/L = 0.66$), and Gt--Hb ($x/L = 0.75$), as we move from anterior to posterior.   In each crossing region, it is plausible that the dynamics of the network are dominated by the interactions among just the pair of genes whose expression levels are crossing.

We argue that criticality in a system of two mutually repressive genes generates several clear, experimentally observable signatures.  First,  there should be nearly perfect anti--correlations between the fluctuations in the two expression levels.  As a result, there are two linear combinations of the expression levels, or ``modes,'' that have very different variances.  Second, fluctuations in the large variance mode should have a significantly non--Gaussian distributions, while the small variance mode is nearly Gaussian.  Third, there should be  a dramatic slowing down of the dynamics along one direction in the space of possible expression levels. Finally,  there should be correlations among fluctuations at distant points in the embryo.  These signatures are related: the small variance mode will be the direction of fast dynamics, and under some conditions the large variance mode will be the direction of slow dynamics; the fast fluctuations should be nearly Gaussian, while the slow modes are non--Gaussian; and only the slow mode should exhibit long--ranged spatial correlations. We will see that all of these effects are found in the gap gene network.  Importantly, these signatures do not depend  on the molecular details.

To see that signatures of criticality are quite general, we consider a broad class of models for a genetic regulatory circuit.    The rate at which gene products are synthesized depends on the concentration of all the relevant transcription factors, and we also expect that the gene products are degraded.  To simplify, we ignore delays, so that the rate at which the protein encoded by a gene is synthesized depends instantaneously on the other protein (transcription factor) concentrations,  and we assume that degradation obeys first order kinetics.  We also focus on a single cell, leaving aside (for the moment) the role of diffusion.  Then, by choosing our units correctly we can write the dynamics for the expression levels of two interacting genes as
 \begin{eqnarray}
\tau_1 {{dg_1}\over{dt}} &=& f_1 (c; g_1 , g_2) - g_1 + \xi_1\label{model1}\\
\tau_2 {{dg_2}\over{dt}} &=& f_2 (c; g_1 , g_2) - g_2 + \xi_2 ,\label{model2}
\end{eqnarray}
where $g_1$ and $g_2$ are the normalized expression levels of the two genes,  $\tau_1$ and $\tau_2$ are the lifetimes of the proteins, and $c$ represents the external (maternal) inputs.  The functions $f_1$ and $f_2$ are the ``regulation functions'' that express how the transcriptional activity of each gene depends on the expression level of all the other genes; with our choice of units, the regulation function runs between zero (gene off) and one (full induction).  All of the molecular details of transcriptional regulation are hidden in the precise form of these regulation functions  \cite{bintu_a}, which we will not need to specify.  Finally, the random functions $\xi_1$ and $\xi_2$ model the effects of noise in the system.  

If the interactions are weak, then for any value of the external inputs $c$ there is a single steady state response, defined by expression levels $\bar g_1 (c)$ and $\bar g_2 (c)$.  We can check whether this hypothesis is consistent by asking what happens to small changes in the expression levels around this steady state.  We write $g_1 = \bar g_1 + \delta g_1$, and similarly for $g_1$, and then expand Eqs (\ref{model1}, \ref{model2}) assuming that $\delta g_1$ and $\delta g_2$ are  small.  The result is
\begin{equation}
{d\over{dt}} 
\left[ \begin{array}{c} 
\delta g_1 \\ 
\delta g_2 
\end{array} \right] 
= 
\left[ \begin{array}{cc} 
-\Gamma_1 & \gamma_{12}\\
\gamma_{21} & -\Gamma_2
\end{array}\right] 
\left[ \begin{array}{c} 
\delta g_1 \\ 
\delta g_2 
\end{array} \right]
+
\left[ \begin{array}{c} 
\eta_1 \\ 
\eta_2 
\end{array} \right] .\label{linear1}
\end{equation}
Here $\Gamma_1$ and $\Gamma_2$ are effective decay rates for the two proteins, which must be positive if the steady state we have identified is stable.  The parameter $\gamma_{12}$ reflects the incremental effect of gene $2$ on gene $1$---$\gamma_{12} < 0$ means that the protein encoded by gene 2 is a repressor of gene $1$---and similarly for $\gamma_{21}$.  The noise terms $\eta_1$ and $\eta_2$ play the same role as $\xi_1$ and $\xi_2$, but have different normalization.

If the steady state that we have identified is stable, then the matrix
\begin{equation}
\hat M \equiv \left[ \begin{array}{cc} 
-\Gamma_1 & \gamma_{12}\\
\gamma_{21} & -\Gamma_2
\end{array}\right] 
\end{equation}
must have two eigenvalues with negative real parts.  This is guaranteed if the interactions are weak ($\gamma_{12}, \gamma_{21} \rightarrow 0$), but as the interactions become stronger it is possible for one of the eigenvalues to vanish.  This is the critical point.  Notice that we can define the critical point without giving a microscopic description of all the interactions that determine the form of the regulation functions.

The linearized Eqs (\ref{linear1}) predict that the relaxation of average expression levels to their steady states can be written as combinations of two exponential decays, 
\begin{equation}
\left[\begin{array}{c}
\langle g_1 (t) \rangle \\ \langle g_2 (t) \rangle
\end{array}\right]
=  
\left[\begin{array}{c}
\bar g_1 (c) \\ \bar g_2 (c)  
\end{array}\right]
+ \left[\begin{array}{cc}
A_{1s} & A_{1f}\\
A_{2s} & A_{2f}\end{array}\right]
\left[\begin{array}{c}
e^{\Lambda_s t} \\  e^{\Lambda_f t} \end{array}\right]
\label{lambdas}
\end{equation}
where $\Lambda_s$ and $\Lambda_f$ are the ``slow'' and ``fast'' eigenvalues of $\hat M$.
Thus, while we measure the two expression levels, there are linear combinations of these expression levels---different directions in the ($g_1, g_2$) plane---that provide more natural coordinates for the dynamics, such that motion along each direction  is a single exponential function of time.   As we approach criticality, the dynamics along the slow direction becomes very slow, so that $\Lambda_s \rightarrow 0$.

The linearized Eqs (\ref{linear1}) also predict the fluctuations around the steady state.   As we approach criticality,  things simplify, and we find the covariance matrix
\begin{equation}
\left[
\begin{array}{cc}
\langle (\delta g_1 )^2\rangle & \langle \delta g_1 \delta g_2\rangle\\
\langle \delta g_1 \delta g_2\rangle & \langle (\delta g_2)^2\rangle
\end{array}
\right]
\rightarrow
\sigma^2 
\left[
\begin{array}{cc}
1  & \Gamma_{1}/\gamma_{12}\\
 \Gamma_{1}/\gamma_{12} &  (\Gamma_{1}/\gamma_{12})^2
\end{array}
\right] ,
\label{Cmatrix}
\end{equation}
where $\sigma^2$ is the variance in the expression level of the first gene.  As with the dynamics, there are two ``natural'' directions in the ($g_1, g_2$) plane corresponding to eigenvectors of this covariance matrix (principal components).   In this linear approximation, the critical point is the point where we ``lose'' one of the dimensions, and the fluctuations in the two expression levels become perfectly correlated or anti--correlated.  In addition, the direction with small fluctuations is the direction of fast relaxation.

Testing the predictions of criticality requires measuring the time dependence of gap gene expression levels, with an accuracy better than the intrinsic noise levels of the system.  Absent live movies of the expression levels,  the progress of cellularization provides a clock that can be used to mark the time during nuclear cycle fourteen at which an embryo was fixed \cite{lecuit+al_02}, accurate to within one minute  \cite{expts}.  Fixed embryos, with immunofluorescent staining of the relevant proteins, thus provide a sequence of snapshots that can be placed accurately along the time axis of development.  Immunofluorescent staining itself provides a measurement of relative protein concentrations that is accurate to within $\sim 3\%$ of the maximum expression levels in the embryo \cite{expts}. 

\begin{figure}[tb]
\includegraphics[width=\linewidth]{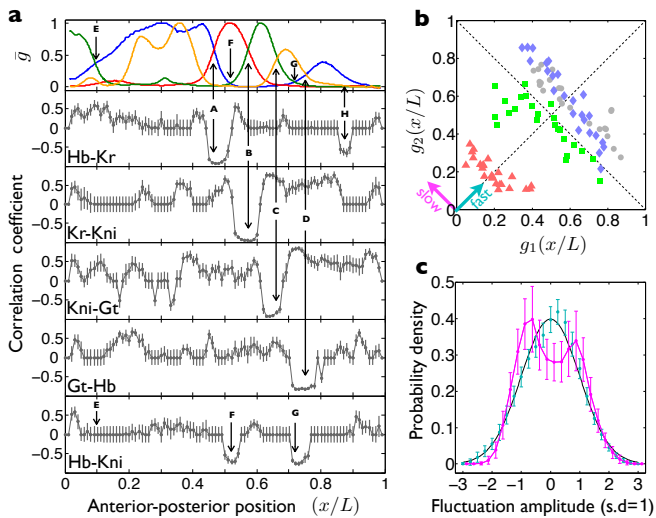}
\caption{Fluctuations in gap gene expression levels. (a) Pairwise correlation coefficients between fluctuations in the different gap genes  vs. anterior--posterior position, from the same data as in Fig \ref{fig1} \cite{expts}.  Mean expression levels at top to guide the eye, colors as in Fig \ref{fig1};  error bars are from bootstrap analysis.  Correlations  which are not significant at $p=0.01$ are shown as zero.  Major crossing points of the mean expression profiles are labelled A, B, C, and D; other points  marked as described in the text.
(b) Scatter plot of expression levels for pairs of genes in individual embryos: Kr vs Hb at point A (grey circles), Kni vs Kr at point B (blue diamonds), Gt vs Kni  at point C (green squares), and Hb vs Gt at point D (red triangles).   (c) Probability distribution of expression fluctuations. In each of the crossing regions from Fig 1,  we form the combinations $\delta g_f$ (fast modes, cyan) and $\delta g_s$ (slow modes, magenta), and normalize the fluctuations across embryos to have unit variance at each position.  Data from all four regions are pooled to estimate the distributions; error bars are from random divisions of the set of 24 embryos.    Gaussian distribution (black) shown for comparison. \label{fig2}}
\end{figure}

 In Fig \ref{fig2}a we show the correlations between fluctuations in pairs of gap genes at each position.  Gap gene expression levels plateau at $\sim 40\,{\rm min}$ into nuclear cycle fourteen \cite{expts}, and the mean expression levels are shown as a function of anterior--posterior position in Fig \ref{fig1}.  At each position we can look across the many embryos in our sample, and analyze the fluctuations around the mean, as in Ref \cite{bits}.    We see that, precisely in the ``crossing region'' where Hb and Kr are the only genes with significant expression (marked A in Fig \ref{fig2}a), the correlation coefficient approaches $C = -1$, perfect (anti--)correlation, as expected at criticality.    This pattern repeats at the crossing between Kr and Kni (B), at the crossing between Kni and Gt (C), and, perhaps less perfectly, at the crossing between Gt and Hb (D) \cite{imperfect}.  These strong anti--correlations are shown explicitly in Fig \ref{fig2}b, where we plot the two relevant gene expression levels against one another at each crossing point.   In all cases, the direction of small fluctuations is along the positive diagonal, while the large fluctuations are along the negative diagonal. 
  
It is important that the strong anti--correlations  tell us something about the underlying network, rather than being a necessary (perhaps even artifactual) corollary of the mean expression profiles.    A  notable feature of Fig \ref{fig2}a thus is what happens away from the major crossing points.  There is a Hb--Kni crossing at $x/L = 0.1$ (E), but this  does not have any signature in the correlations, perhaps because spatial variations in expression levels at this point are dominated by maternal inputs  rather than being intrinsic to the gap gene network  \cite{torso1,torso2}.   This is evidence that we can have crossings without correlations, and we can also have correlations without crossings, as with Hb and Kr at point H; interestingly, H marks the point where an additional posterior Kr stripe appears during gastrulation \cite{2ndKr1,2ndKr2}.  We also note that strong correlations can appear when expression levels are very small, as with
Hb and Kni  at points F and G;  there also are  extended regions of positive Kr--Kni and Kni--Gt correlations in parts of the embryo where the expression levels of Kr and Kni both are very low.  Taken together, these data indicate that the pattern of correlations is not simply a reflection of the mean spatial profiles, but an independent measure of network behavior. 

If we transfrom Eq (\ref{linear1})  to a description in terms of the fast and slow modes $g_f$ and $g_s$, then precisely at criticality there is no ``restoring force'' for fluctuations in $g_s$ and formally the variance $\sigma^2$ in Eq (\ref{Cmatrix}) should diverge.  This is cut off by higher order terms in the expansion of the regulation functions around the steady state, and this leads to a non--Gaussian distribution of fluctuations in $g_s$.   Although the data are limited, we do find, as shown in Fig \ref{fig2}c,  that fluctuations in the small variance (fast) direction are almost perfectly Gaussian, while the large variance (slow) direction show  significant departures form Gaussianity, in the expected direction.

\begin{figure}[tb]
\includegraphics[width=\linewidth]{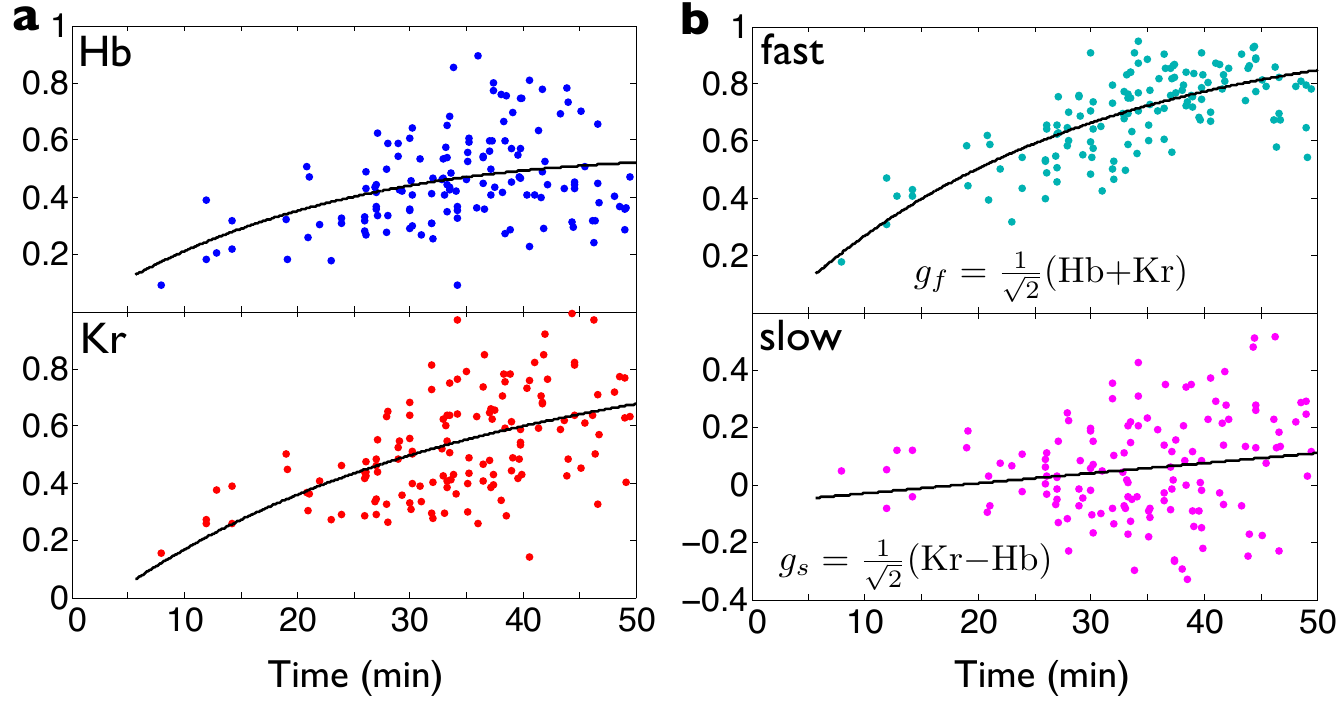}
\caption{Dynamics at the Hb--Kr crossing point.  (a) Normalized expression levels of the individual genes, plotted vs. time during nuclear cycle fourteen; data from Ref \cite{expts}.   (b) Linear combinations of expression levels corresponding to the small variance (fast) and large variance (slow) directions in Fig \ref{fig2}b.  Curves are best fit single exponentials for each mode and are also shown projected back into the individual expression levels in (a).  Eigenvalues, as in Eq (\ref{lambdas}), are $\Lambda_f = -0.04 \pm 0.01\, {\rm min}^{-1}$ and $\Lambda_s = -0.002 \pm 0.007\, {\rm min}^{-1}$.  \label{fig3}}
\end{figure}

 The time dependence of Hb and Kr expression levels during nuclear cycle fourteen is shown, at the crossing point $x/L = 0.47$, in Fig \ref{fig3}.  Criticality predicts that if we take a linear combination of these expression levels corresponding to the direction of small fluctuations in Fig \ref{fig2}b (cyan), then we will see relatively fast dynamics, and this is what we observe.  In contrast, if we project onto the direction of large fluctuations (magenta), we see only very slow variations over nearly one hour.  Indeed, the expression level along this slow direction seems almost to diffuse freely, with growing variance rather than systematic evolution.  Thus, strong (anti--)correlations are accompanied by a dramatic slowing of the dynamics along one direction in the space of possible expression levels, and a similar pattern is found at each of the crossings, Kr--Kni, Kni--Gt, and Gt--Hb (data not shown).  Again, this is consistent with what we expect for two--gene systems at criticality.

If we move along the anterior-posterior axis in the vicinity of the crossing point, the sum of expression levels of two genes, which is proportional to the fast mode, remains approximately constant, while the difference, which is proportional to the slow mode, changes. Therefore the dynamics of the slow mode, shown in Fig \ref{fig3}, will generate motion of the pattern along the anterior--posterior axis. This slow shift is well known \cite{expts,jaeger+al_04}.

The slow dynamics associated with criticality also implies that correlations should extend over long distance in space.  As is clear from Fig \ref{fig3}, the eigenvalues $\Lambda_s$ and $\Lambda_f$ define time scales $\tau_s = -1/\Lambda_s$ and $\tau_f = -1/\Lambda_f$. If we add diffusion to the dynamics in Eqs (\ref{linear1}), then these time scales define length scales, through the usual relation  $\ell_{s,f} = \sqrt{D\tau_{s,f}}$; although there is some dependence on details of the underlying model, these lengths define the distances over which we expect fluctuations to be correlated.  In particular,  as we approach criticality, $\Lambda_s$  vanishes and the associated correlation length $\ell_s$ can become infinitely long, limited only by the size $L$ of the embryo itself.  Searching for these long--ranged correlations is complicated by the fact that the system is inhomogeneous, but we have a built in control, since we should see the long ranged correlations only in the slow, large variance mode $\delta g_s$, and not in $\delta g_f$.  This control also helps us discriminate against systematic errors that might have generated spurious correlations.

\begin{figure}[b]
\includegraphics[width=\linewidth]{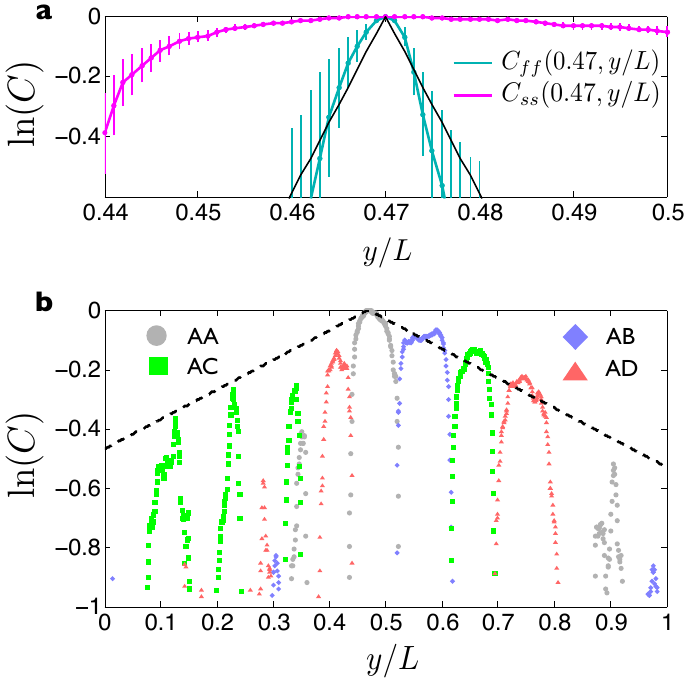}
\caption{   Spatial correlations of fluctuations in gene expression.  (a) Auto--correlations of slow (magenta) and fast (cyan) modes near the Hb--Kr crossing, as defined in Eq (\ref{Css}).  Line shows a fit, $C_{ff} (x,y) \propto e^{-|x-y|/\xi}$, with $\xi/L = 0.015 \pm 0.002$.   (b) Correlations between the slow mode at the Hb--Kr crossing and slow modes at other crossings, as defined in Eq (\ref{CssAB}), indicated by the colors.  All correlation functions are evaluated at $x/L=0.47$ with $y$ varying as shown.  Middle peak is the same (auto--)correlation function $C_{ss}$ as in (a). Dashed line is $e^{-|x-y|/\xi}$, with $\xi = L$.  \label{fig4}}
\end{figure}

In Fig \ref{fig4}a we show the normalized correlation function
\begin{equation}
C_{ss}(x,y) = {{\langle \delta g_s (x) \delta g_s (y)\rangle}\over{\left(\langle [\delta g_s (x)]^2\rangle\langle[ \delta g_s (y)]^2\rangle\right)^{1/2}}} ,
\label{Css}
\end{equation}
with $x$ held fixed at the Hb--Kr crossing and $y$ allowed to vary.  We see that this correlation function is essentially constant throughout the crossing region.  In contrast, the same correlation computed for the fast mode decays rapidly, with a length constant $\xi/L\sim 0.02$, just a few nuclear spacings  along the anterior--posterior axis.  

The dominant slow mode corresponds to different combinations of expression levels in different regions of the embryo.  Generally, we can write the slow mode as a weighted sum of the different expression levels,
\begin{equation}
g_s(x) = \sum_{{\rm i} =1}^4 W_{\rm i} (x) g_{\rm i}(x),
\end{equation}
where we label the gap genes ${\rm i}=1$ for Hb, ${\rm i}=2$ for Kr, ${\rm i}=3$ for Kni, and ${\rm i}=4$ for Gt.  Near the Hb--Kr crossing, labelled A in Fig \ref{fig2}a, we have $W_{\rm i} \approx W_{\rm i}^A$, where $W_{\rm i}^A$ are the weights that give us the anti--symmetric combination of Hb and Kr, as drawn in Fig \ref{fig2}b: $W^A_1 = -1/\sqrt{2}$, $W^A_2 = 1/\sqrt{2}$, and $W^A_3 = W^A_4 = 0$.  Similarly, near the Kr--Kni crossing, labelled B in Fig \ref{fig2}a, we have $W_{\rm i} \approx W_{\rm i}^B$ with $W^B_2 = -1/\sqrt{2}$, $W^B_3 = 1/\sqrt{2}$, and $W^B_1 = W^B_4 = 0$, and this generalizes to crossing regions C and D.  Using these weights, we obtain approximations to the slow mode,
\begin{equation}
\left[\begin{array}{c}
g_s^A(y)\\ g_s^B(y)\\  g_s^C(y) \\  g_s^D(y)
\end{array}\right]
=
\sum_{{\rm i}=1}^4 
\left[\begin{array}{c}W_{\rm i}^A g_{\rm i}(y) \\
 W_{\rm i}^B g_{\rm i}(y) \\
 W_{\rm i}^C g_{\rm i}(y) \\
 W_{\rm i}^D g_{\rm i}(y) \end{array}\right],
 \label{ABCD}
\end{equation}
and we expect that these approximations are accurate in their respective crossing regions.  Now we can test for correlations over longer distances by computing, for example,
\begin{equation}
C_{ss}^{AB}(x,y) = {{\langle \delta g_s^A (x) \delta g_s^B (y)\rangle}\over{\left(\langle [\delta g_s^A (x)]^2\rangle\langle[ \delta g_s^B (y)]^2\rangle\right)^{1/2}}} ,
\label{CssAB}
\end{equation}
holding $x/L = 0.47$ in the crossing region $A$ while letting $y$ vary through the crossing region $B$, and similarly for  $C_{ss}^{AC}(x,y)$ and $C_{ss}^{AD}(x,y)$.  The results of this analysis are shown in Fig \ref{fig4}b; note that $C_{ss}^{AA}(x,y)$ is the correlation we have plotted in Fig \ref{fig4}a.

Figure \ref{fig4}b shows that the slow mode is correlated over very long distances.  We can see, for example, in $C_{ss}^{AC}$, correlations between fluctuations in expression level at the Hb--Kr crossing region and at the Kni--Gt crossing region, despite the fact that these regions are separated by $\sim 20\%$ of the length of the embryo and have no significantly expressed genes in common.  These peaks in the correlation functions appear also at points anterior to the crossing regions, presumably at places where our approximations in Eq (\ref{ABCD}) come close to some underlying slow mode in the network.  The pattern of correlations has an envelope  corresponding to an exponential decay with correlation length $\xi =L$ (dashed line in Fig \ref{fig4}b), and similar results are obtained for the correlation functions $C_{ss}^{BC}$, $C_{ss}^{BD}$, $C_{ss}^{CD}$, etc..  This means that fluctuations in expression level are correlated along essentially the entire length of the embryo, as expected at criticality.

To summarize,   the patterns of gap gene expression in the early {\em Drosophila} embryo exhibit several signatures of criticality:  near perfect anti--correlations of fluctuations in the expression levels of different genes at the same point, non--Gaussian distributions of the fluctuations in the large variance modes, slowing down of the dynamics of these  modes, and  spatial correlations of the slow modes that extend over a large fraction of the embryo.  While each of these observations could have other explanations, the confluence of results strikes us as highly suggestive.  Note that we have focused on aspects of the data that are connected to the hypothesis of criticality in a very general way, independent of other assumptions, rather than trying to build a model for the entire network.

The possibility that biological systems might be poised near critical points, often discussed in the past \cite{bak}, has been re--invigorated by new data and analyses on systems ranging from ensembles of amino acid sequences to networks of neurons to flocks of birds \cite{mora+bialek_11}.  In the specific context of transcriptional regulation, the approach to criticality serves to generate long time scales, which may serve to reduce noise and optimize information transmission \cite{tkacik+al_12a}.  For the embryo in particular, these long time scales and the corresponding long length scales may give us a different view of the problem of scaling expression patterns to variations in the size of the egg \cite{scaling,bialek_12}.

Even leaving aside the possibility of criticality, the aspects of the data that we have described here are not at all what we would see if the gap gene network is described by generic parameter values. There must be something about the system that is finely tuned in order to generate such large differences in the time scales for variation along different dimensions in the space of expression levels, or to insure that correlations are so nearly perfect and extend over such long distances.

\begin{acknowledgments}
We thank S  Blythe, K Doubrovinski, O Grimm, JJ Hopfield, S Little, M Osterfield, AM  Polyakov, M  Tikhonov, G Tka\v{c}ik, and A Walczak for helpful discussions.  We are especially grateful to EF Wieschaus, for discussions both of the ideas and their presentation.  This work was supported in part by NSF Grants PHY--0957573  and CCF--0939370, by NIH Grants P50 GM071508  and R01 GM097275, by the Howard Hughes Medical Institute,  by the WM Keck Foundation, and by Searle Scholar Award 10--SSP--274 to TG.  
\end{acknowledgments}


\begin{thebibliography}{99}
\bibitem{imperfect}
Since we observe substantial anti--correlations both in the Gt--Hb pair and in the  Hb--Kni pair, it is likely that the system is more nearly three dimensional in the neighborhood of this crossing, so that no single pair can achieve perfect correlation.
\end{thebibliography}
\end{document}